\documentclass[reprint, amsmath, amssymb, aps, prl]{revtex4-2}

\usepackage{graphicx}
\usepackage{dcolumn}
\usepackage{bm}
\usepackage{hyperref}
\usepackage{physics}
\usepackage{xfrac}
\usepackage{xcolor}
\usepackage{soul}

\newcommand{\comment}[1]{}

\begin{document}

\preprint{APS/123-QED}

\title{A low-loss and broadband all-fiber acousto-optic circulator}

\author{Martin Blaha}
\author{Arno Rauschenbeutel}
\author{Riccardo Pennetta}
\email{riccardo.pennetta@hu-berlin.de}

\affiliation{%
 Department of Physics, Humboldt Universit\"at zu Berlin, 10099 Berlin, Germany\\
}%

\date{\today}

\begin{abstract}
The introduction of low-loss optical fibers probably represents the single most important advance in the growth of our telecommunication system. To meet our needs for secure communications, it is likely that our classical network will soon be operating alongside what is known as a quantum network. The latter is very sensitive to loss and thus poses new constraints to the performance of current fiber components. In particular, recent quantum network prototypes underlined the absence of low-loss non-reciprocal fiber-based devices. 
Here, we present a solution to this issue by realizing low-loss (0.81 dB), broadband (at least 50 GHz bandwidth) and high-extinction (up to 27 dB) circulators, based on Mach-Zehnder interferometers including so-called fiber null-couplers. The latter are directional couplers, whose splitting-ratio can be controlled by launching acoustic waves along the coupling region. Fabricated from standard single-mode fibers and actuated electrically, these circulators can be made to fit any existing optical fiber networks and could turn out to be key for the transmission and processing of optically encoded quantum information.

\end{abstract}

\maketitle

\subsection{Introduction}

Non-reciprocal devices, such as isolators and circulators, constitute an important part of many optical setups and find application in simple tasks, such as protecting a laser from back-reflections to more demanding operations as signal routing and processing in optical networks. Most off-the-shelf devices rely on the Faraday effect, an approach that is not easy to implement in integrated devices, whether they are based on fiber-optics or photonic circuits. Here, the most prominent challenges lie in fabrication complexity, large material losses and difficulty in handling magnetic fields. For instance, commercially available fiber-coupled circulators rely on free-space optics and feature high insertion-loss of about 1 dB at telecom wavelengths. This insertion loss increases even significantly for near infrared and visible light, i.e., the wavelength regions in which most quantum emitters operate. Currently, the lack of low-loss fiber-coupled circulators constitutes a substantial obstacle for the development of fiber-based quantum networks \cite{Hacker2016, Niemietz2021, Witthaut2012, Duan2005}, for which loss represents a highly undesired additional source of decoherence~\cite{Reiserer2015}.

For these reasons, extensive effort has been devoted to identify alternative ways to break Lorentz reciprocity, as for instance using non-linear optics \cite{Bino2018, White2022}, electro-optics \cite{Lira2012}, chirally coupled atoms \cite{Scheucher2016, Pucher2022}, and acousto-optics. The latter has been particularly successful because of the strong coupling between light and acoustic vibration that can be achieved when both waves are confined to small volumes, as for instance in tapered \cite{Birks1996} and photonic-crystal fibers \citep{Dainese2006, Kang2011} or in photonic circuits \cite{Kittlaus2021, Sohn2021, Tian2021, Zhou2024} and whispering gallery mode resonators \cite{Kim2017}. From a practical perspective, this approach is very promising for the next generation of fiber-based non-reciprocal devices as it potentially allows low insertion loss, broadband operations and electrical actuation.

Here, we report the experimental demonstration of a novel type of all-fiber acousto-optic circulator, realized by cascading two so-called fiber null-couplers \cite{Birks1996} to form a Mach-Zehnder interferometer. These  circulators are electrically actuated and feature low insertion loss (0.81 dB), high extinction ratio up to 27 dB, and a broad bandwidth of at least 50 GHz. Furthermore, being fabricated using standard single-mode fiber, they can be seamlessly integrated in any existing or future fiber network.

\subsection{Fiber null-couplers}

\begin{figure*}[]
\includegraphics[width=1\linewidth]{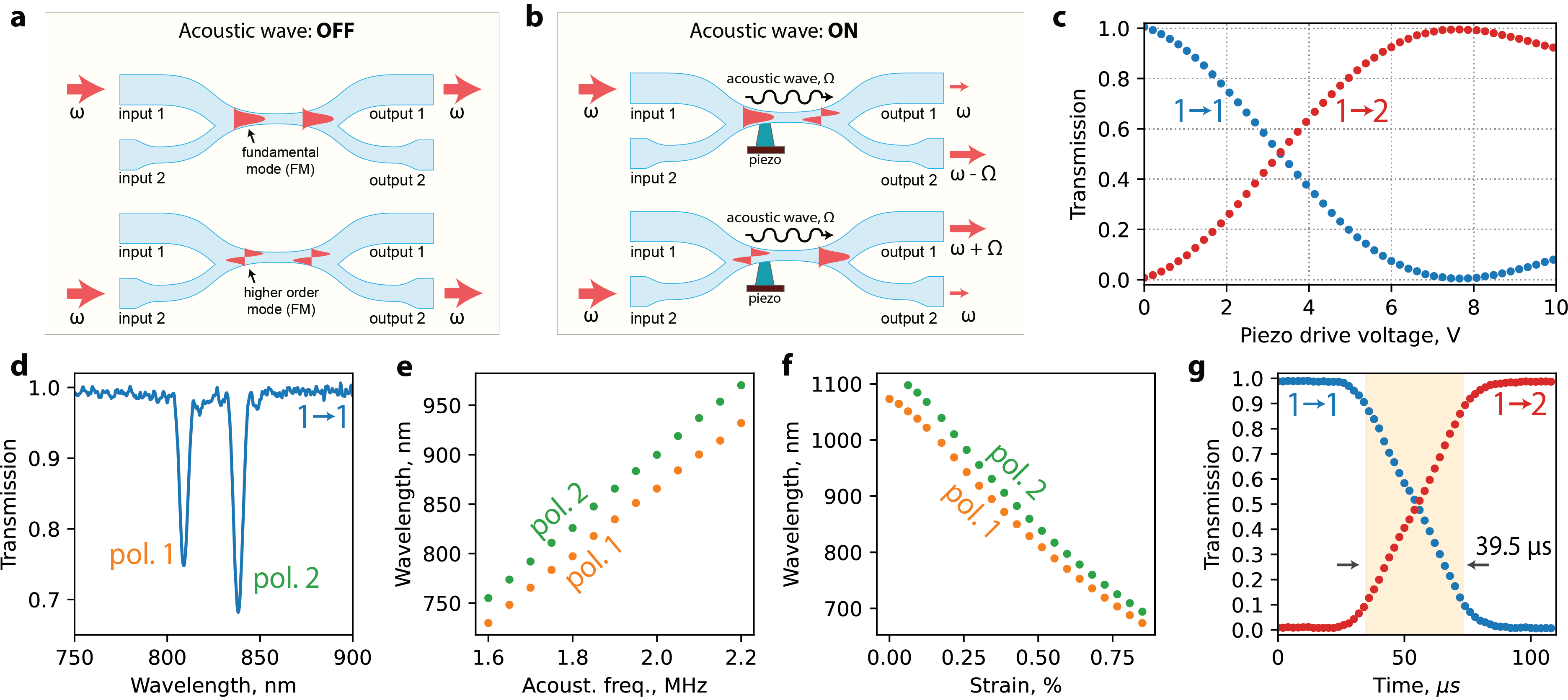}
\caption{\label{fig:FigNC} 
(a, b) Schematic representation of the working principle of a null-coupler.
(c) Measured transmission of a null-coupler as a function of the peak-to-peak amplitude of the sinusoidal voltage signal applied to the piezo-actuator (acoustic frequency $\Omega$ = 2.0 MHz). The blue and red dots refer to the transmission measured from input 1 to output 1 and from input 1 to output 2, respectively.
(d) Measured null-coupler transmission using a spectrometer after excitation with a white light source. The two observed peaks correspond to phase-matching for two orthogonal polarization states.
(e) Measured phase-matching wavelength as a function of the acoustic frequency, $\Omega$.
(f) Measured phase-matching wavelength as a function of the strain applied to the null-coupler for a a fixed acoustic frequency of $\Omega$ = $2 \pi$ 2.0 MHz.
(g) Measured response time of a null-coupler after sudden excitation of the piezo-actuator. The 10:90 rise/fall-time is marked by a yellow-shaded area and was found to be $39.5\mu s$.
}
\end{figure*}

The elementary unit of our circulators is a particular type of fiber coupler known as null-coupler \cite{Birks1996, Farwell1998}. A standard fiber coupler is fabricated by heating and fusing together two identical fibers. Controlling the coupler length and degree of fusion allows one to obtain any coupling ratio between $0\%$ and $100\%$. However, if the fibers have different diameters, the maximum achievable coupling is less than $100\%$. A null-coupler is a fiber coupler made with fibers so dissimilar that the maximum coupling is approximately zero. In this case, adiabatic propagation along the coupler ensures that light launched into the larger fiber excites the fundamental mode (FM) of the coupler waist, eventually emerging from the same fiber at the other end of the coupler. The same occurs for light launched into the thinner fiber, which instead excites the first higher order mode (HOM) of the coupler waist (see Fig.~\Ref{fig:FigNC}~(a)).
Acousto-optic interaction can transform a null-coupler in an arbitrary splitting-ratio coupler. Let us consider the configuration in Fig.~\Ref{fig:FigNC}(b), in which a piezo-electric transducer is employed to launch an acoustic wave along the coupler waist, causing a periodic modulation of its refractive index. If phase-matching occurs, i.e., if the acoustic wavelength matches the beat-length between the FM and the HOM, the acoustic wave can mediate intermodal coupling of light between these two modes. The coupling strength and, consequently, the amount of coupled light can be controlled by changing the amplitude of the launched acoustic wave (see Fig.~\ref{fig:FigNC}(a,b)).
Figure~\Ref{fig:FigNC}(c) shows a measurement of this process for a null-coupler fabricated in our self-built fiber post-processing setup using standard Corning HI780C optical fiber. All the measurements reported in the following were performed at a wavelength of 764 nm, unless otherwise stated. The insertion loss of the coupler was 0.18 dB and the maximum passive splitting ratio was $0.04\%$, i.e., the device performed well as a null-coupler. Upon driving the piezo-actuator with an electric signal with a frequency of 2.0~MHz, we found that up to $99.5\%$ of the transmitted optical power could be coupled across the device. It should be noted that most of the experiments described below are carried out with an acoustic frequency of 2.0~MHz, because the available piezo-electric actuators have a mechanical resonance, which makes them perform particularly well around this value. We would also like to underline that, in general, null-couplers have very low electrical power requirement, meaning that they can be driven by low cost, general purpose function generators.

A simple method of finding the phase matching condition in a null-coupler consists in exciting its guided mode using a white light source and measuring its transmission with a spectrometer. Since most null-couplers exhibit optical birefringence, this measurement usually reveals two distinct peaks, which correspond to phase-matching for two orthogonal polarization states of the launched light (see Fig.~\Ref{fig:FigNC}(d)). For a fixed null-coupler geometry and driving acoustic frequency $\Omega$, the phase-matching condition is fulfilled over an optical bandwidth of a few nanometers (approximately 5 nm for the device shown in Fig.~\Ref{fig:FigNC}(d)). By varying $\Omega$, phase-matching can be achieved for any optical wavelength over the entire working bandwidth of the selected optical fiber type (see Fig.~\Ref{fig:FigNC}(e)). Remarkably, for a fixed null-coupler geometry and a fixed acoustic frequency, it is still possible to very widely tune the phase-matching condition by straining the null-coupler waist, as depicted in Fig.~\Ref{fig:FigNC}(f). These different methods to precisely tune the phase-matching condition make null-couplers very versatile, allowing one to compensate possible fabrication tolerances and, e.g., easily fabricate several null-couplers operating at the same optical wavelength and acoustic frequency.
Finally, we note that the response time of the null-couplers is given by the time required for acoustic wave front to travel along the coupler waist and usually corresponds to few tens of $\mu s$ (see Fig.~\Ref{fig:FigNC}(g)).

\subsection{Acousto-optic non-reciprocity and working principle}

\begin{figure*}[]
\includegraphics[width=1\linewidth]{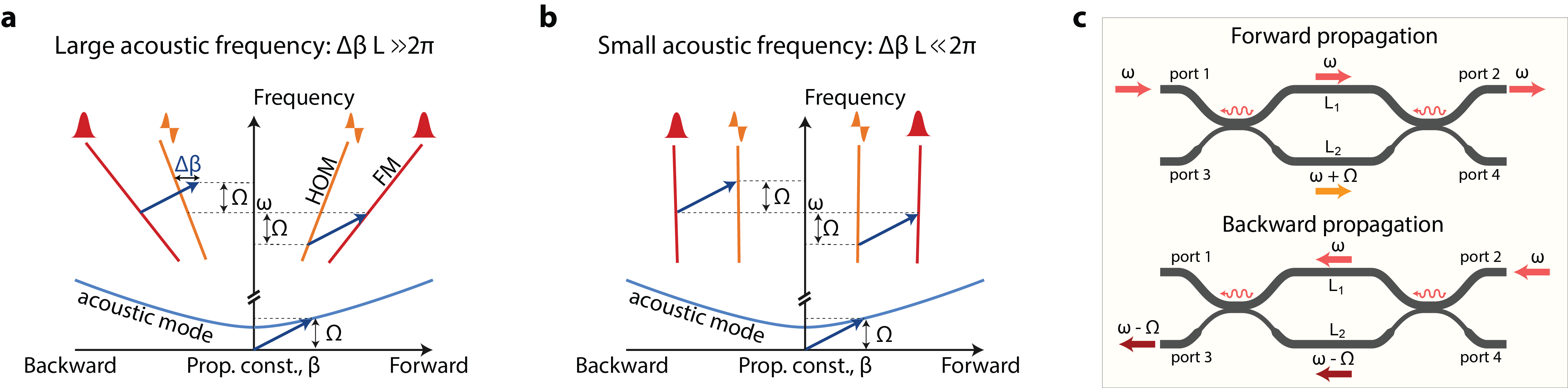}
\caption{\label{fig:FigTheo} 
(a,b) Dispersion diagrams for optoacoustic  intermodal coupling: $\omega$ indicates the light frequency, $\beta$ its propagation constant and $\Omega$ is the acoustic frequency.
(a) For large acoustic frequencies, phase-matches only occurs in the forward direction. The mismatch of the propagation constant in the backward direction is indicated with $\Delta\beta$. (b) In state-of-the-art electrically actuated all-fiber devices, the achievable mechanical frequencies are usually small, therefore the same acoustic can mediate coupling both in forward and backward direction. The resulting device is, however, still not reciprocal because the sign of the frequency shift experienced by the propagating light depends on its propagation direction.
(c) Schematic of the proposed null-coupler-based circulator, indicating the forward (upper panel) and backward (lower panel) light propagation. Due to acousto-optical interaction in the null-couplers, the light in the lower arm of the interferometer is subject to a propagation direction-dependent frequency shift. This enables the use of the interferometer as a 4-port circulator, if the system parameters satisfy the condition $\Delta\varphi_{NR} = \pi$. 
}
\end{figure*}

Launching travelling acoustic waves along a waveguide breaks the symmetry between forward and backward light propagation and, as a result, acousto-optic intermodal coupling is non-reciprocal. To understand in detail why this is the case, let us consider the dispersion diagram in Fig.~\Ref{fig:FigTheo}(a), in which the frequency $\omega$ of the propagating optical mode is shown as a function its propagation constant $\beta$ for the FM and the HOM. Let us now assume that an acoustic wave also propagates along the waveguide, whose frequency and momentum perfectly phase-match intermodal coupling between the FM and the HOM in the forward direction. Such an acoustic wave, in general, does not phase-match intermodal coupling for backward propagating light fields (Fig.~\Ref{fig:FigTheo}(b)). In particular, this occurs when the total phase mismatch: $\Delta \beta L = (\beta_{FM} + \beta_{HOM}) L \Omega / \omega \gg 2\pi$ \cite{Kittlaus2018}. This is the key physical idea behind many of the acousto-optic non-reciprocal devices reported in the literature (see, e.g., Refs.~\cite{Yu2009a, Kang2011, Kittlaus2018, Kittlaus2021, Sohn2021, Tian2021, Kim2017}). Note that, since usually $L \le 10$ cm, these systems had to be designed so that $\Omega \gg 1$ GHz.

For fiber-based devices, the excitation of such high-frequency acoustic waves is challenging and, although this has been demonstrated using auxiliary guided laser fields \cite{Kang2011}, the resulting devices were unsuited for further applications due to the high optical power required. From a practical point of view, electrical actuation is far more convenient. However, to the best of our knowledge, no technique has been demonstrated yet that enables the electrical excitation of GHz acoustic waves propagating along optical fibers.
In fact, null-couplers work at lower acoustic frequencies, $\Omega$, in the MHz range and, therefore, their dispersion is more accurately described by the diagram in Fig.~\ref{fig:FigTheo}(b). In this case, the same acoustic wave can mediate intermodal coupling in both the forward and backward. Hence, a single null-coupler does not have non-reciprocal re-routing capabilities. Nonetheless, the light that is coupled across a null-coupler via acousto-optic interaction is still subject to a non-reciprocal frequency shift, whose sign depends on the relative direction of propagation of the light and acoustic wave (see Fig.~\Ref{fig:FigTheo}(b)). This equals $-\Omega$ if the light and acoustic wave are co-propagating and $+\Omega$ in the opposite case.
To use this to our advantage, let us consider the interferometer shown in Fig.~\ref{fig:FigTheo}(c), where two null-couplers are cascaded in a Mach-Zehnder configuration. Both null-couplers are driven at the same acoustic frequency and both coupling ratios are set to  $50 \%$. As depicted in Fig.~\ref{fig:FigTheo}(c), when light is launched from port 1, acousto-optic interaction in the first null-couplers shifts the frequency of the light field propagating in the lower arm of the interferometer by $+\Omega$. This shift is later compensated at the second null-coupler, such that the light at port 2 exhibits no frequency shift and its intensity is given by: $ I_{1 \rightarrow 2} = \frac{1}{2} I_0  [ 1 + \textrm{cos}( \Delta \varphi_{1 \rightarrow 2})]$. Here, the relative propagation phase, $\Delta\varphi_{1\rightarrow 2}$ is given by:
\begin{equation}
\Delta \varphi_{1 \rightarrow 2} = \frac{n \omega}{c} \Delta L - \frac{n \Omega}{c} L_2 \ ,
\label{Eq:Forward}
\end{equation}
where, $I_0$ is the intensity of the input light and $\Delta L = L_1 - L_2$, where $L_1$ and $L_2$ are the lengths of the two arms of the interferometers. Instead, if light is launched from port 2, light propagating in the lower arm of the interferometer experiences a frequency shift of $–\Omega$ and the intensity of the light reaching port 1 is given by: $ I_{1 \leftarrow 2} = \frac{1}{2} I_0  [ 1 + \textrm{cos}( \Delta \varphi_{1 \leftarrow 2})]$, where:
\begin{equation}
\Delta \varphi_{1 \leftarrow 2} = \frac{n \omega}{c} \Delta L + \frac{n \Omega}{c} L_2.
\label{Eq:Backward}
\end{equation}
Comparing Eqs.~\ref{Eq:Forward} and~\ref{Eq:Backward} we derive the non-reciprocal phase shift, $\Delta \varphi_{NR}$:
\begin{equation}
\Delta \varphi_{NR}=\Delta \varphi_{1 \leftarrow 2}-\Delta \varphi_{1 \rightarrow 2}=2 \frac{n \Omega}{c} L_2
\label{eq:NonRec}
\end{equation}
An optical circulator can be realized by satisfying the condition: $\Delta\varphi_{NR} = \pi$. In this case, operating the interferometer such that light launched from port 1 constructively interferes at port 2 ensures that light back-reflected from port 2 is completely re-routed to port 3 (see Fig.~\ref{fig:FigTheo}(c)). This mechanism also works when light is launched from ports 2, 3 and 4, i.e., the proposed device is a 4-port circulator. Furthermore, in a Mach-Zehnder interferometer, the intensity at each of the output ports is independent of the frequency of the input light if the lengths of the two arms are equal: $L_1 = L_2$ (see Eqs. (1, 2)). Thus, satisfying this second condition ensures broadband operations, only limited by the bandwidth of null-couplers, that exceeds several hundred GHz.

\begin{figure}[]
\includegraphics[width=1\linewidth]{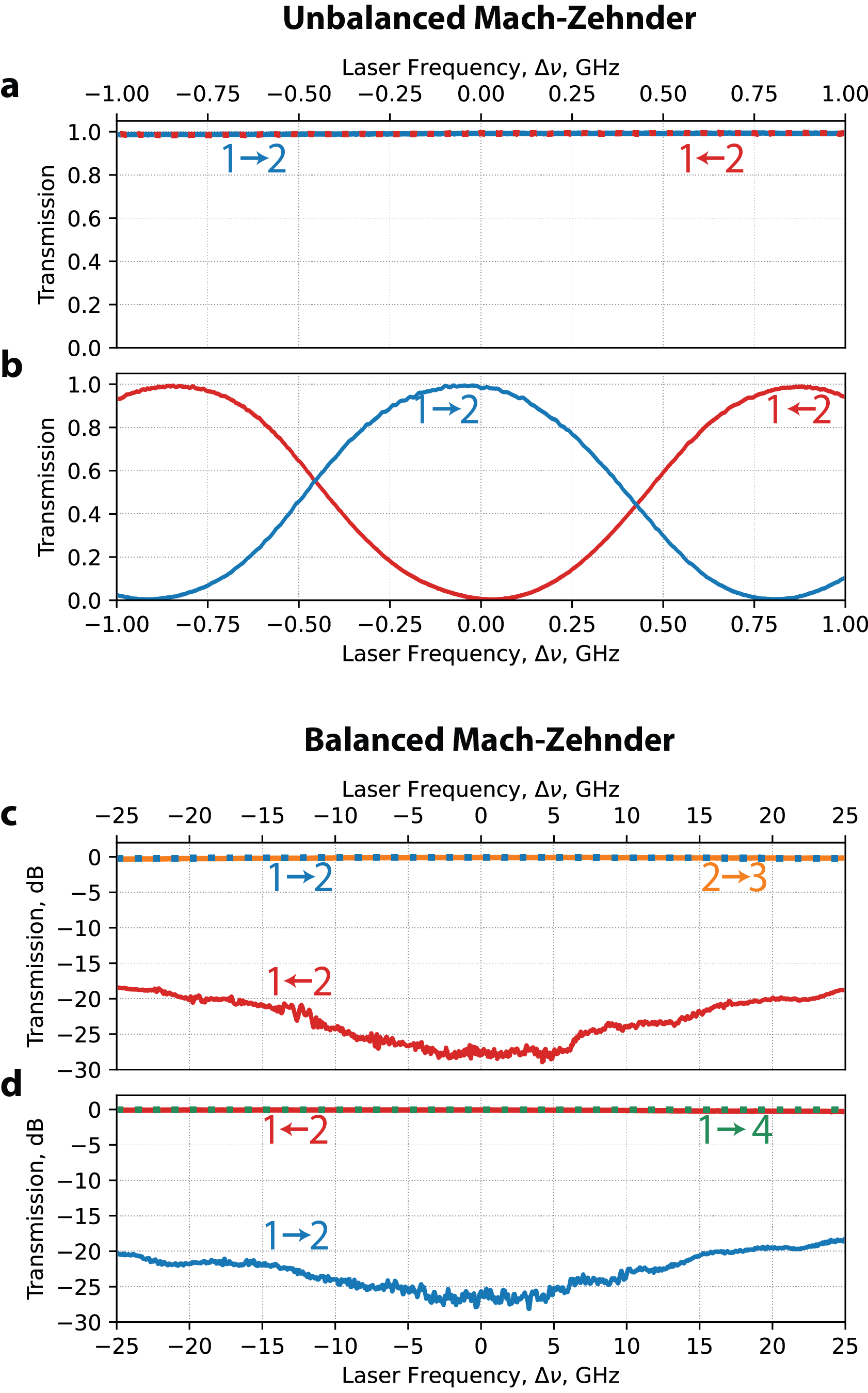}
\caption{\label{fig:FigCirc}
(a, b) Transmission properties of the proposed circulator for the unbalanced ($\Delta L = 12$ cm) Mach-Zehnder configuration. (a) Measured transmission as a function of the laser frequency, $\Delta \nu$, in the $1 \rightarrow 2$ (blue solid line) and $1 \leftarrow 2$ (dashed red line), when both null-couplers are off. (b) Same measurement, with both null-couplers switched on. In this case transmission in the $1 \leftarrow 2$ direction is indicated by a solid red line.
(c, d) Transmission properties in the case of the balanced ($\Delta L = 190 \ \mu$m) Mach-Zehnder configuration. (c) Measured transmission as a function of the laser frequency, $\Delta \nu$, in the $1 \rightarrow 2$ (blue solid line), $1 \leftarrow 2$ (solid red line) and $3 \leftarrow 3$ (dashed orange line) direction. (d) Same measurement, but after changing the working point of the interferometer to have high transmission in the $1 \leftarrow 2$ (solid red line) and $1 \rightarrow 4$ (dashed green line) direction and high extinction ratio in the $1 \rightarrow 2$ direction.
}
\end{figure}

\subsection{Experimental demonstration of a null-coupler circulator}

To experimentally demonstrate these ideas, we have assembled a Mach-Zehnder interferometer using two null-couplers operating at the same mechanical frequency of 2.0 MHz. In this case, from Eq.~\Ref{eq:NonRec}, non-reciprocal light propagation requires $L_2 = 25.8$ m. In order to precisely satisfy the condition $L_1 = L_2$, this length of fiber was spooled around two home-made fiber stretchers, which allowed us to adjust the optical path length of each of the two arms of the interferometers over a maximum range of a few cm. In addition, two in-line polarization controllers were placed in each arm of the interferometer to tune the polarization state of the propagating light field at the null-couplers. Again, all measurements were performed at a wavelength of 764 nm.

Due to the limited frequency scanning range of our laser source (approximately 50 GHz), to verify the working mechanism of the circulator, we started by running the interferometer in an unbalanced configuration (i.e., the two arms had different lengths). Under this condition, and we measured the transmission from port 1 to port 2 and vice-versa as a function of the laser frequency, $\Delta \nu$ (see Fig.~\Ref{fig:FigCirc}). With the null-couplers off, the optical response was fully reciprocal and, as expected, no interference fringes were visible (Fig.~\Ref{fig:FigCirc}(a)). When, switching on both null-couplers, high-visibility interference fringes appeared that exhibit a clearly non-reciprocal behaviour (Fig.~\Ref{fig:FigCirc}(b)). Around $\Delta \nu = 0$ constructive (destructive) interference was observed in the $1 \rightarrow 2$ ($1 \leftarrow 2$) direction, i.e., light was allowed to propagate in a single direction. We note that the distance between two minima in the intereference fringes is given by $\Delta \nu_\text{min} = \frac{c}{n \Delta L}$. Using this relation together with Eq.~\ref{eq:NonRec}, we estimated $\Delta L$ = 12 cm and $\Delta \varphi_{NR} = 1.06 \  \pi$, by fitting the experimental data. 

To explore the full potential of the circulator, we then balanced the interferometer by cutting a fiber length of $\approx 10$ cm from the longer of the two arms and then finely matching the respective optical path lengths using the fiber stretchers. We then performed transmission measurements when launching light in the $1 \rightarrow 2$ and $1 \leftarrow 2$ directions, as depicted in Fig.~\Ref{fig:FigCirc}(c). An extinction-ratio as high as 27 dB was measured in this configuration. Optical circulation was demonstrated by verifying that light propagating in the $1 \leftarrow 2$ direction was then re-routed to port 3, as indicated in Fig.~\Ref{fig:FigCirc}(c) (transmission in the $3 \leftarrow 2$ direction was not shown in Fig.~\Ref{fig:FigCirc}(a,b) for clarity). By balancing the interferometer, broadband operations of the circulator were obtained, indeed, we measured an extinction ratio of at least 18 dB over the entire 50-GHz frequency scan of our laser. By fitting the experimental data, we estimated $\Delta L$ = 290 $\mu$m and a corresponding -10 dB bandwidth of $\approx 140$ GHz. Further reducing the value of $\Delta L$ was difficult given the available frequency scanning range of the laser. To demonstrate that the circulator is a true 4-port device, we repeated these measurements by adjusting the working point of the interferometer so that light was allowed to propagate in the $1 \leftarrow 2$ direction, while light in the $1 \rightarrow 2$ direction was completely re-routed to port 4.  As shown in Fig.~\Ref{fig:FigCirc}(d), very similar performance were obtained in this case. The fabricated device exhibits insertion loss of 0.81 dB, which has been measured via cut-back. Of these, 0.40 dB are due to the insertion loss of the two couplers, 0.12 dB to propagation loss in the optical fibers in the two arms of the interferometer and the remaining 0.29 dB are attributed to fiber splices (two in each interferometer arm) and additional optical components (i.e. two polarization controllers in each arm).

\subsection{Discussion}

The demonstrated circulators have a number of very attractive features. First, even in this proof-of-principle demonstration, the insertion loss are comparable to or lower than fiber-coupled circulators currently available from major fiber component suppliers, to the best of our knowledge. We would like to emphasize that our measurements were performed in a wavelength range that is of high interest for quantum information processing, but where the availability of high-quality fiber components is considerably lower than, e.g., at telecom wavelengths. In addition, they are true 4-port devices, whereas most commercially devices are only 3-port devices.
Compared to other acousto-optic platforms, such as photonic integrated circuits, null-coupler-based devices rely on a much simpler fabrication technique, have lower electrical power requirements, and offer tremendous flexibility. In fact, each fabricated device can be tuned to operate over a wide range of mechanical frequencies and over the entire operating range of the selected fiber type by adjusting the strain along the coupler waist. Moreover, they can be readily integrated in any fiber network with low-loss via fusion splicing.
Finally, the demonstrated bandwidth is already well beyond the requirements of experiments involving quantum emitters such as laser-cooled atoms \cite{Niemietz2021, Lechner2023, Reiserer2015}, single molecules \cite{Faez2014} and vacancies \cite{OrphalKobin2023}.

Regarding current limitations, perhaps the most important is that, here, optical circulation has been demonstrated for a single input polarization, meaning that these devices would not be suitable for experiments featuring polarization-encoded qubits. However, this is not a fundamental limitation. Indeed, null-couplers with polarization-independent operations have already been demonstrated \cite{Farwell1998} by controlling the degree of fusion of the two fibers allowing the modal birefringence to be perfectly compensated by the geometrical birefringence. Thus, circulators built with such couplers would work for arbitrary polarization states.
Another practical issue is that, as with other Mach-Zehnder based optical devices (e.g. electro-optic amplitude modulators), the working point of the interferometer must be actively stabilized if operation over long time scales is desired. This can be implemented by including a piezo-driven fiber stretcher in one of the arms of the interferometer \cite{Lechner2023}.
Finally, a possible drawback is that light leaving the device after being subjected to optical circulation undergoes a frequency shift, which may be undesirable for certain applications (see Fig.~\ref{fig:FigTheo}(c)). This, can however be solved by introducing a third null-coupler, which would then return the light frequency to its original value.

\subsection{Conclusions}

In summary, we have shown that by cascading two null-couplers in a Mach-Zehnder configuration it is possible to realize all-fiber circulators with excellent performance. Beyond the intrinsic interest of demonstrating acousto-optic non-reciprocity using low acoustic frequencies, we believe that, given their compact size and relatively simple and low-cost fabrication procedure, not dissimilar from the one of standard fiber couplers, the proposed devices could become a widely used tool for many experiments in optics and photonics and set new standards in terms of low-loss in optical circulation.

During the final stages of manuscript preparation, we became aware of a paper exploiting a similar idea in an integrated photonic platform \cite{Cheng2024}.


\begin{acknowledgments}
We thank Thomas Hoinkes, Philipp Schneeweiss and J\"urgen Volz for helpful discussions. We acknowledge financial support by the Alexander von Humboldt Foundation in the framework of an Alexander von Humboldt Professorship endowed by the Federal Ministry of Education.
\end{acknowledgments}

\bibliography{main.bib}

\newpage
\appendix

\end{document}